\documentclass[12pt]{article}
\usepackage{amsthm,amssymb,amsmath}
\usepackage[english]{babel}

 1 
1

\renewcommand{\d}{\operatorname{d}}

\newcommand{\be}{\begin{equation}}
\newcommand{\ee}{\end{equation}}
\usepackage[dvips]{graphicx}
\begin{document}
\title{\sc Exact Solutions
of Integrable 2D Contour Dynamics
\thanks{Partially supported by DGCYT
project BFM2002-01607 }}
\author{Luis Mart\'{\i}nez Alonso$^{1,\ddag}$
 and  Elena Medina$^{2,\maltese}$\\
\emph{ $^1$Departamento de F\'{\i}sica Te\'{o}rica II, Universidad
Complutense}\\ \emph{E28040 Madrid, Spain} \\
\emph{$^2$Departamento de Matem\'{a}ticas, Universidad de
C\'{a}diz}\\\emph{ E11510, Puerto Real, C\'{a}diz, Spain}\\
\texttt{$^\ddag$luism@fis.ucm.es}\\
\texttt{$^{\maltese}$elena.medina@uca.es}}
\date{} \maketitle
\begin{abstract}
A class of exact solutions of the dispersionless Toda hierarchy
constrained by a string equation is obtained. These solutions
represent deformations of analytic curves with a finite number
of nonzero harmonic moments. The corresponding $\tau$-functions
are determined and the emergence of cusps is studied.

\end{abstract}

\vspace*{.5cm}

\begin{center}\begin{minipage}{12cm}

\end{minipage}
\end{center}
\newpage

\section{Introduction}

Integrable contour dynamics governed by the dispersionless Toda
(dToda) hierarchy is a multifaceted subject. It underlies
problems of complex analysis \cite{1},\cite{2}, interface
dynamics (Laplacian growth) \cite{3} , Quantum Hall effect
\cite{4} and associativity (WDVV) equations \cite{5}. A common
ingredient in many of its applications is the presence of
random models of normal $N\times N$ matrices \cite{1}-\cite{4},
\cite{6},\cite{7} with partition functions of the form
\begin{equation}\label{1}
Z_N=\int \d M\d M^{\dagger}
\exp(-\frac{1}{\hbar}\,\mbox{tr}\,\,W(M,M^{\dagger})),
\end{equation}
where
\begin{equation}\label{2}
W(z,\bar{z})=z\bar{z}+v_0-\sum_{k\geq 1}(t_k\,z^k+\bar{t}_k\,\bar{z}^k).
\end{equation}
In an appropriate large $N$ limit ($\hbar\rightarrow 0$,
$s:=\hbar N$ fixed), the eigenvalues of the matrices are
distributed within a planar domain (\emph{support of eigenvalues})
with sharp edges,
which depends on the parameters $t:=(s=\bar{s},\,t_1,t_2\ldots)$.

 If the support of eigenvalues is
a simply-connected bounded domain with boundary given by an
analytic curve $\gamma\, \,(z=z(p),\, |p|=1)$,  then
$(s,t_1,t_2\ldots)$ are harmonic moments of $\gamma$ and the curve
evolves with $(t,\bar{t})$ according to the dToda hierarchy.
Moreover, the corresponding $\tau$-function represents the
quasiclassical limit of the partition function \eqref{1}. A
particularly interesting feature is that for almost all initial
conditions the evolution of $\gamma$ leads to critical
configurations in which cusp-like singularities develop. This
behaviour is well-known in Laplacian growth \cite{8} and random
matrix theory \cite{9}.

In order to obtain solutions of the dToda hierarchy describing
contour dynamics one must impose a string equation which
leads to a particular type of Riemann-Hilbert problem
\cite{10}-\cite{12}. In this paper we present a method for
finding solutions in the form of  Laurent polynomials
\begin{equation}\label{3}
z=rp+u_0+ \cdots+\frac{u_{K-1}}{p^{K-1}},
\end{equation}
which describe dynamics of curves with a finite number of nonzero
harmonic moments, namely $t_k=\bar{t}_k=0$ for $k\geq K$. We
exhibit examples for arbitrary $K$ and derive  their
corresponding $\tau$-functions. Furthermore the  emergence of
cusps is analytically studied.

\section{dToda contour dynamics}

Let $z=z(p)$ be an invertible conformal map of the exterior of the
unit circle to the exterior of a simply connected domain bounded
by a simple analytic curve $\gamma$ of the form
\begin{equation}\label{4}
\bar{z}=S(z),
\end{equation}
where bar stands for complex conjugation ($z(\bar{p})=z(p^{-1})$
on $\gamma$) and the \emph{Schwarz function} $S(z)$ is analytic in
some domain containing $\gamma$.

 The map $z(p)$ can be represented by
a Laurent series
\begin{equation}\label{5}
z(p)=r p+\sum_{k=0}^{\infty}\frac{u_k}{p^k},
\end{equation}
with a real coefficient $r$. The coefficients
$(r,u_0,u_1,\ldots)$ are functions of the harmonic moments
$t=(s=\bar{s},\,t_1,t_2\ldots)$ of the exterior of $\gamma$,
which in turn can be introduced through the expansion of the
Schwarz function
\begin{equation}\label{6}
S(z)=\sum_{k=1}^{\infty}k\,t_k\,z^{k-1}+\frac{s}{z}+\sum_{k=1}^
{\infty}\frac{v_k}{z^{k+1}},
\end{equation}
with $(v_1,v_2,\ldots)$ being  functions dependent on $t$. As a
consequence of \eqref{1}-\eqref{3}, it follows that
$z(p,t,\bar{t})$ solves the dToda hierarchy
\begin{align}\label{7}
\nonumber
\partial_{t_k}z&=\{H_k,z\},\quad \partial_{\bar{t}_k}z=-\{\bar{H}_k,z\},\\\\
\nonumber H_k:&=(z^k)_{\geq 1}+\frac{1}{2}(z^k)_0,\quad
\bar{H}_k:=(\bar{z}^k)_{\leq -1}+\frac{1}{2}(\bar{z}^k)_0.
\end{align}
where $\{f,g\}:=p\,(\partial_p\,f\,\partial_s\,g
-\partial_p\,g\,\partial_s\,f),$ the function $\bar{z}(p^{-1})$
is defined by the Laurent series
\begin{equation}\label{8}
\bar{z}(p^{-1}):=\frac{r}{p}+\sum_{k=0}^{\infty}\bar{u}_k\,p^k,
\end{equation}
and the symbols $(\ldots)_{\geq 1}$ ($(\ldots)_{\leq -1}$) and
$(\ldots)_0$ mean truncated Laurent series with only positive
(negative) terms and the constant term, respectively. Furthermore,
this solution satisfies the string equation

\begin{equation}\label{9}
\{z(p),\bar{z}(p^{-1})\}=1.
\end{equation}

These properties can be proved through the twistor scheme of
Takasaki-Takebe \cite{3}. It uses  Orlov-Schulman functions of the
dToda hierarchy
\begin{align}\label{10}
\nonumber m&=\sum_{k=1}^{\infty}k\,t_k\,z^k+s+\sum_{k=1}^
{\infty}\frac{v_k}{z^k},\\\\
\nonumber \bar{m}&=\sum_{k=1}^{\infty}k\,\bar{t}_k\,\bar{z}^k+
s+\sum_{k=1}^ {\infty}\frac{\bar{v}_k}{\bar{z}\,^k},
\end{align}
and can be summarized as follows:

\vspace{0.3truecm} \noindent {\bf Theorem}\emph{  If
$(z,m,\bar{z}, \bar{m})$ are functions of $(p,t,\bar{t})$ which
admit expansions of the form \eqref{5},\eqref{8},\eqref{10} and
satisfy the equations
\begin{equation}\label{11}
\bar{z}=\frac{m}{z},\quad \bar{m}=m,
\end{equation}
then $(z,\bar{z})$ is a solution of the dToda hierarchy
constrained by the string equation \eqref{9}. }

\section{Solutions}

Equations \eqref{11} are meaningful only when they are interpreted
as a suitable Riemann-Hilbert problem on the complex plane of the
variable $p$. Thus $(z,m)$  must be analytic functions in a
neighborhood $D=\{|p|>r\}$ of $p=\infty$  and $(\bar{z},\bar{m})$
must be analytic functions in a neighborhood $D'=\{|p|<r'\}$ of
$p=0$. The statement of the Theorem holds provided $A:=D\bigcap
D'\neq \emptyset$.

We next prove that equations \eqref{11} have solutions satisfying
\eqref{5},\eqref{8} and \eqref{10} with
\[
t_k=  0,\quad  k> K,\quad t_{K}\neq 0.
\]
In this way we assume
\begin{equation}\label{12} \everymath{\displaystyle}
\begin{array}{rcl}
m&=&\sum_{k=1}^{K}nt_k z^k+s+\sum_{k=1}^{\infty}\frac{v_k}
{z^k},\\  \\
\bar{m}&=&\sum_{k=1}^{K}k\,\bar{t}_k \bar{z}^k+s+
\sum_{k=1}^{\infty}\frac{\bar{v}_k}{\bar{z}^k}.
\end{array}
\end{equation}

Given two integers $r_1\leq r_2$ we denote by
$\mathrm{V}[r_1,r_2]$ the set of Laurent polynomials of the form
\[
c_{r_1}\,p^{r_1}+c_{r_1+1}\,p^{r_1+1}+\cdots+c_{r_2}\,p^{r_2}.
\]

Let us look for solutions of \eqref{11} such that $z$ and
$\bar{z}$ are meromorphic functions of $p$ with possible poles at
$p=0$ and $p=\infty$ only. Then, as a consequence of the
assumptions \eqref{5},\eqref{8} and \eqref{12}, from \eqref{11} it
follows that
\begin{equation}\label{13}
z\in\mathrm{V}[1-K,1],\quad \bar{z}\in\mathrm{V}[-1,K-1].
\end{equation}
The equation $\bar{m}=m$ is equivalent to the system:
\begin{equation}\label{14}
\bar{m}_{\geq1}=m_{\geq 1},
\end{equation}
\begin{equation}\label{15}
\bar{m}_0=m_0,
\end{equation}
\begin{equation} \label{16}
\bar{m}_{\leq -1}=m_{\leq -1}.\end{equation}
If we now set
\begin{equation}\label{17}
 m=\bar{m}=\sum_{k=1}^{K}nt_k(z^k)_{\geq1}+
\bar{m}_0 +\sum_{k=1}^{K}k\,\bar{t}_k (\bar{z}^k)_{\leq -1},
\end{equation}
with
\[
\bar{m}_0=s+\sum_{k=1}^{K}k\,\bar{t}_k (\bar{z}^k)_0.
\]
it can be easily seen  that $\bar{m}$ has the required expansion
of the form \eqref{12} provided $z$ and $\bar{z}$ satisfy
\eqref{5} and \eqref{8} . On the other hand, the expression
\eqref{17} for $m$ has an expansion of the form \eqref{12} if the
residue of $\displaystyle\frac{m}{z}$
 corresponding to its Laurent expansion in powers of
$z$ verifies

\begin{equation}\label{18}
Res(\frac{m}{z},z)=s.
\end{equation}
Hence the problem reduces  to finding $z$ and $\bar{z}$ satisfying
\eqref{5},\eqref{8},\eqref{18} and
\begin{equation}\label{19}
z=\frac{m}{\bar{z}}.
\end{equation}
In view of \eqref{13} we look for $z$ and $\bar{z}$ of the form
\begin{align} \label{20}
\nonumber &z=rp+u_0+
\cdots+\frac{u_{K-1}}{p^{K-1}},\\\\
\nonumber &\bar{z}=\frac{r}{p}+\bar{u}_0+
\cdots+\bar{u}_{K-1}p^{K-1}.
\end{align}
Now, in order to prevent  $z$ from having poles
different from $p=0$ and $p=\infty$ we have to impose
\begin{equation}\label{21}
m(p_i)=0,
\end{equation}
where $p_i$ denote the $K$ zeros of
\[
r+\bar{u}_0\,p+ \cdots+\bar{u}_{K-1}\,p^K=0.
\]
In this way by using the expression \eqref{17} of $m$, the only
variables appearing in \eqref{19} are
\[
(p,t,\bar{t},r,u_0,\ldots,u_{K-1},w_0, \ldots,w_{K-1}),\quad
w_i:=\bar{u}_i.
\]
Thus, by identifying coefficients of the powers $p^i,\,
i=1-K,\ldots,1$ we get $K+1$ equations which together with the $K$
equations \eqref{21} determine the  $2K+1$   unknowns variables
$(r,u_0,\ldots,u_{K-1},w_0, \ldots,w_{K-1})$ as functions of
$(t,\bar{t})$. Moreover, provided $r$ is a real coefficient, the
equations \eqref{11} are invariant under the transformation
\[
\mathrm{T}f(p)=\overline{f\Big(\frac{1}{\bar{p}}\Big)}.
\]
Hence if $(r,u_0,\ldots,u_{K-1},w_0, \ldots,w_{K-1})$ solves
\eqref{19} so does
\[
(r,\bar{w}_0,\ldots,\bar{w}_{K-1},\bar{u}_0,\ldots,\bar{u}_{K-1}).
\]
Therefore, if both solutions are close
enough, they coincide and consequently $w_i=\bar{u}_i$, as required.

To complete our proof we must show that \eqref{18} is satisfied
too. To do that let us take two circles $\gamma\; (|p|=r)$ and
$\gamma '\; (|p|=r')$ in the complex $p$-plane  and denote by
$\Gamma$ and $\Gamma '$ their images under the maps $z=z(p)$ and
$\bar{z}=\bar{z}(1/p)$, respectively. Notice that due to \eqref{5}
and \eqref{8}, the curves $\Gamma$ and $\Gamma '$ have positive
orientation if $\gamma$ and $\gamma '$ have positive and negative
orientation, respectively. Then we have
\begin{align*}
&Res(\frac{m}{z},z)-Res(\frac{\bar{m}}{\bar{z}},\bar{z})=\frac{1}{2i\pi}\oint_{\Gamma} \frac{m}{z}\d z-
\frac{1}{2i\pi}\oint_{\Gamma '} \frac{\bar{m}}{\bar{z}}
\d \bar{z}\\\\
&=\frac{1}{2i\pi}\oint_{\gamma} \bar{z}\partial_p z\,\d p-
\frac{1}{2i\pi}\oint_{\gamma '} z\partial_p\bar{z}\,\d p=\frac{1}{2i\pi}\oint_{\gamma} \partial_p(\bar{z}z)\,\d p=0,
\end{align*}
where we have taken into account that the integrands are analytic
functions of $p$ in $\mathbb{C}-\{0\}$ and that $\gamma$ and the
opposite curve of $\gamma'$ are homotopic with respect to
$\mathbb{C}-\{0\}$.  Therefore, as we have already proved that
$\bar{m}$ has an expansion of the form \eqref{12}, we deduce
\[
Res(\frac{m}{z},z)=Res(\frac{\bar{m}}{\bar{z}},\bar{z})=s,
\]
so that \eqref{18} follows.

Let us  illustrate the method with the case $K=2$. The polynomial
$p\,\bar{z}$ has two zeros at the points
$$p_1={\frac{-\bar{u}_0 + {\sqrt{{{\bar{u}_0}^2} -
         4\,r\,\bar{u}_1}}}{2\,\bar{u}_1}},
         \quad
p_2={\frac{-\bar{u}_0 - {\sqrt{{{\bar{u}_0}^2} -
         4\,r\,\bar{u}_1}}}{2\,\bar{u}_1}},$$
and from \eqref{21} we get two equations which lead to
\begin{equation}\label{22}\everymath{\displaystyle}
\begin{array}{l}
-2\,{r^2}\,t_2\,{{\bar{u}_0}^3} +
   4\,{r^3}\,t_2\,\bar{u}_0\,\bar{u}_1 +
   r\,t_1\,{{\bar{u}_0}^2}\,\bar{u}_1 +
   4\,r\,t_2\,u_0\,{{\bar{u}_0}^2}\,
    \bar{u}_1 - {r^2}\,t_1\,{{\bar{u}_1}^2} -
   4\,{r^2}\,t_2\,u_0\,{{\bar{u}_1}^2}\\  \\
    -
   s\,\bar{u}_0\,{{\bar{u}_1}^2} -
   \bar{t}_1\,{{\bar{u}_0}^2}\,{{\bar{u}_1}^2} -
   2\,\bar{t}_2\,{{\bar{u}_0}^3}\,{{\bar{u}_1}^2} +
   r\,\bar{t}_1\,{{\bar{u}_1}^3} = 0,
\end{array}
\end{equation}
\begin{equation}\label{23}\everymath{\displaystyle}
\begin{array}{l}
-2\,{r^3}\,t_2\,{{\bar{u}_0}^2} +
   2\,{r^4}\,t_2\,\bar{u}_1 +
   {r^2}\,t_1\,\bar{u}_0\,\bar{u}_1 +
   4\,{r^2}\,t_2\,u_0\,\bar{u}_0\,
    \bar{u}_1 - r\,s\,{{\bar{u}_1}^2} -   r\,\bar{t}_1\,\bar{u}_0\,{{\bar{u}_1}^2}
    \\  \\
     -   2\,r\,\bar{t}_2\,{{\bar{u}_0}^2}\,{{\bar{u}_1}^2} -
   2\,{r^2}\,\bar{t}_2\,{{\bar{u}_1}^3} = 0.
\end{array}
\end{equation}
Identification of  the powers of $p$ in \eqref{19} implies
\begin{equation}\label{24}\everymath{\displaystyle}\begin{array}{lcl}
p:&  &-2\,{r^2}\,t_2 + r\,\bar{u}_1 = 0,\\  \\
p^0:&  & 2\,{r^2}\,t_2\,\bar{u}_0 -
   r\,t_1\,\bar{u}_1 -
   4\,r\,t_2\,u_0\,\bar{u}_1 +
   u_0\,{{\bar{u}_1}^2} = 0,\\  \\
p^{-1}:&  &
-2\,{r^2}\,t_2\,{{\bar{u}_0}^2} +
   2\,{r^3}\,t_2\,\bar{u}_1 +
   r\,t_1\,\bar{u}_0\,\bar{u}_1 +
   4\,r\,t_2\,u_0\,\bar{u}_0\,\bar{u}_1 -
   s\,{{\bar{u}_1}^2} - \bar{t}_1\,\bar{u}_0\,
    {{\bar{u}_1}^2} \\  \\
    &  &- 2\,\bar{t}_2\,{{\bar{u}_0}^2}\,
    {{\bar{u}_1}^2} - 4\,r\,\bar{t}_2\,{{\bar{u}_1}^3} +
   u_1\,{{\bar{u}_1}^3} = 0.
\end{array}\end{equation}
Then by solving equations \eqref{22}-\eqref{24} we get the
solution:
\begin{equation}\label{25}
z=
{\frac{p\,{\sqrt{s}}}{{\sqrt{1 - 4\,t_2\,\bar{t}_2}}}} +
  {\frac{2\,{\sqrt{s}}\,\bar{t}_2}
    {p\,{\sqrt{1 - 4\,t_2\,\bar{t}_2}}}} -
  {\frac{\bar{t}_1 + 2\,t_1\,\bar{t}_2}
    {-1 + 4\,t_2\,\bar{t}_2}},
\end{equation}
which corresponds to the conformal map describing  an \emph{ellipse
growing from a circle} \cite{6}

\subsection*{Solutions for $K\geq 3$ }

Exact solutions associated to arbitrary values of $K$ can be found
from the previous scheme. However in order to avoid complicated
expressions , we set
$$t_1=t_2=\cdots=t_{K-1}=\bar{t}_1=\bar{t}_2=\cdots=\bar{t}_{K-1}=0.$$
and look for particular solutions satisfying
$$u_1=u_2=\cdots=u_{K-2}=\bar{u}_1=\bar{u}_2=\cdots=\bar{u}_{K-2}=0,$$
or equivalently
\begin{equation}\label{26}
z=r\,p+\frac{u_{K-1}}{p^{K-1}},\quad
\bar{z}=\frac{r}{p}+\bar{u}_{K-1}p^{K-1}.
\end{equation}
Under the previous assumptions and from \eqref{17} we have that
\begin{equation}\label{27}
m=Kt_Kr^Kp^K+s+K^2\bar{t}_Kr^{K-1}\bar{u}_{K-1}+\frac{K\bar{t}_Kr^K}{p^K}.
\end{equation}
Thus, we see that \eqref{21} leads us to a unique equation since
from \eqref{27} it follows that $m$ depends on $p$ through $p^K$.
Furthermore,  if $p_i$ satisfies $\bar{z}(p_i)=0$, then
$$p_i^K=-\frac{r}{\bar{u}_{K-1}}.$$ Therefore, \eqref{21} becomes
\begin{equation}\label{28}
s-\frac{Kr^{K+1}t_K}{\bar{u}_{K-1}}+(K-1)Kr^{K-1}\bar{t}_{K}\bar{u}_{K-1}=0.
\end{equation}
On the other hand, it is easy to see that
$$\frac{m}{\bar{z}}=\frac{Kr^Kt_K}{\bar{u}_{K-1}}p+\left(K^2r^{K-1}\bar{t}_K+
\frac{s\,\bar{u}_{K-1}-Kr^{K+1}t_K}{\bar{u}_{K-1}^2}\right)\frac{1}{p^{K-1}},$$
consequently, by equating coefficients and taking \eqref{26} into
account, we find that \eqref{19} leads to  two equations only .
More precisely
\begin{equation}\label{29}\everymath{\displaystyle}\begin{array}{lcl}
p:&   &r=\frac{Kr^Kt_K}{\bar{u}_{K-1}},\\  \\
p^{-(K-1)}&  &
u_{K-1}=K^2r^{K-1}\bar{t}_{K}+\frac{s\,\bar{u}_{K-1}-Kr^{K+1}t_K}{\bar{u}_{K-1}^2}.
\end{array}\end{equation}
Then, we get three equations for the three unknowns $r$,
$u_{K-1}$, $\bar{u}_{K-1}$, which proves that there exits a
solution of the form \eqref{26}. In fact, by solving
\eqref{28}-\eqref{29} we find that
\begin{equation}\label{30}
z=r\,p+\frac{K\,\bar{t}_K\,r^{K-1}}{p^{K-1}},\end{equation} with
$r$ satisfying the implicit equation
\begin{equation}\label{31}
K^2(K-1)t_K\bar{t}_Kr^{2(K-1)}-r^2+s=0.
\end{equation}

Figures 1 and 2 show examples of the evolution of the curve $z(p),\, |p|=1$
as $s$ grows and $t_K$ is kept fixed.

\vspace{5mm}

\begin{figure}[h]
\centering
\includegraphics[width=8cm]{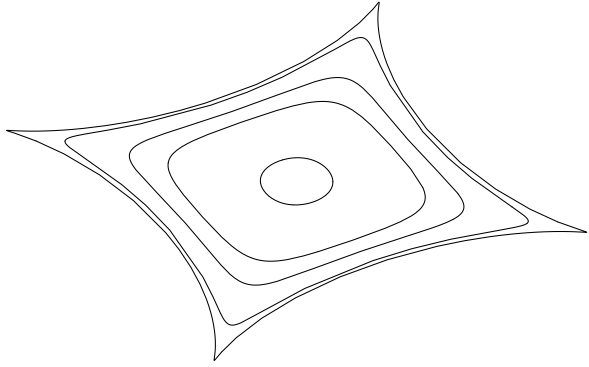}
\caption{solution corresponding to $K=4$}
\end{figure}

\vspace{5mm}

\begin{figure}[h]
\centering
\includegraphics[width=8cm]{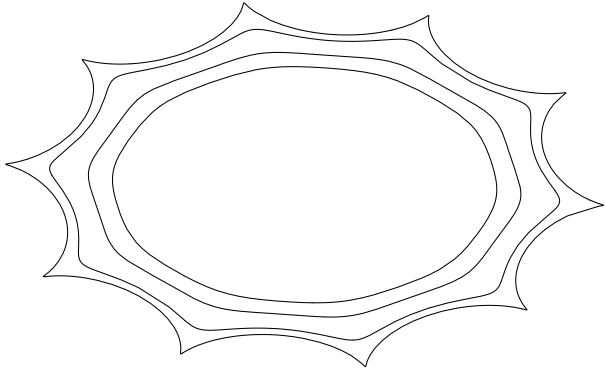}
\caption{solution corresponding to $K=10$}
\end{figure}

\vspace{0.5truecm}

\subsection*{$\tau$-functions }

In \cite{1} it was proved that there is a dToda  $\tau$-function
associated to each analytic curve $z=z(p),\quad |p|=1$, given by
\begin{equation}\label{32}
2\,\log\tau=-\frac{1}{2}s^2+s\,v_0-\frac{1}{2}\sum_{k\geq 1}
(t_k\,v_k+\bar{t}_k\,\bar{v}_k),
\end{equation}
where $v_k$ are the coefficients of the expansion \eqref{6}, and
$v_0$ is determined by
\begin{equation}\label{33}
\frac{\partial v_0}{\partial s}=\log\, r^2,\quad v_0=\frac{\partial \log\,\tau}{\partial s}.
\end{equation}
For the class of solutions \eqref{30} we have
\begin{equation}\label{34}
2\,\log\tau=-\frac{1}{2}s^2+s\,v_0-\frac{1}{2}
(t_K\,v_K+\bar{t}_K\,\bar{v}_K),
\end{equation}
and from \eqref{11} and \eqref{30} it follows that
\begin{align}\label{35}
\nonumber v_K=\frac{1}{2i\pi}\oint_{\Gamma} \bar{z}\,z^K\d z=&
\frac{1}{2i\pi}\oint_{\Gamma} \bar{z}(p) z(p)^K\,
(r-(K-1)\frac{u_{K-1}}{p^{K}})
\d p\\\\
\nonumber =&\frac{(r^2-s)(Ks-(K-2)r^2)} {2K(K-1)s}.
\end{align}
On the other hand by differentiating \eqref{34} with respect to
$s$ and by taking into account \eqref{33} one finds
\begin{align}\label{36}
\nonumber v_0=&-s+s\log\,
r^2+\frac{(K-2)(K-1)K|t_K|^2(Ks-(K-2)r^2)r^{2K}}{2(
(K-1)^2K^2|t_K|^2r^{2K}-r^4)}\\\\
\nonumber +&\frac{(K-2)(s-r^2)}{2(K-1)K}
\Big(K+\frac{(K-2)r^4}{(K-1)^2K^2|t_K|^2r^{2K}-r^4}\Big).
\end{align}
Thus, \eqref{34}-\eqref{36} and \eqref{31} characterize  the
$\tau$-function of the curves determined by \eqref{30}.

\subsection*{Cusps}

The pictures of the curves associated with \eqref{30}-\eqref{31}
show the presence of cusps at some value of $s$ for each fixed
value of $t_K$ . Indeed by using the parametric equation
$p=e^{i\,\theta},\, (0\leq\theta\leq 2\pi)$ for the unit circle,
we have that cusps on the curve $z=z(p)$ appear at points  where
$z_{\theta}=0,\, z_{\theta\theta}\neq 0$ and
$z_{\theta\theta\theta}/z_{\theta\theta}$ has a nonzero imaginary
part. Therefore a necessary condition for $p=p(\theta)$ is
\[
\frac{\partial z}{\partial p}(p)=0,\quad |p|=1.
\]
Thus from \eqref{30} we deduce
\[
p^K=K(K-1)\bar{t}_K\,r^{K-2},
\]
which together with the condition $|p|=1$ requieres that
\begin{equation}\label{37}
r=(K(K-1)|t_K|)^{-\frac{1}{K-2}},
\end{equation}
at some value $s=s(t_K)$. But according to \eqref{31} one finds
that this happens at the value $s_0$ given by
\begin{equation}\label{38}
s_0=\frac{K-2}{K-1}(K(K-1)|t_K|)^{-\frac{2}{K-2}},
\end{equation}
which is the point at which the profile of both positive branches
of $r$, as functions of $s$,  develop an infinite slop (see figure
3).

\begin{figure}[h]
\centering
\includegraphics[width=8cm]{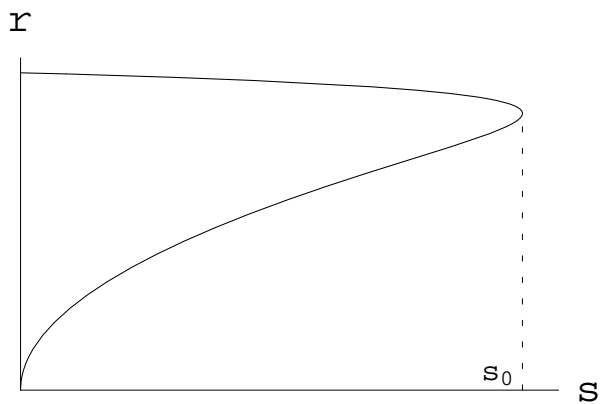}
\caption{The positive branches of $r(s)$ for $K=10$}
\end{figure}
Therefore,  there are $K$ cusps given by the roots
\begin{equation}\label{39}
z_j=\frac{K}{K-1}\Big(\frac{r^2-s_0}{K\,t_K}\Big)^{\frac{1}{K}},
\end{equation}
which emerge when $s$ reaches the extreme value $s_0$ of the
domain of existence of the two positive branches of $r$ as a
function of $s$.

\end{document}